\documentclass[12pt]{iopart}
\usepackage{graphicx}
\begin{document}

\letter{Cu nuclear spin-spin coupling in the dimer singlet state in
SrCu$_{2}$(BO$_{3}$)$_{2}$}

\author{K. Kodama, J. Yamazaki, M. Takigawa\footnote{To whom correspondence
should be addressed (masashi@issp.u-tokyo.ac.jp)}, H. Kageyama, K. Onizuka and Y. Ueda}

\address{Institute for Solid State Physics, Uiversity of Tokyo,
5-1-5 Kashiwanoha, Kashiwa-shi, Chiba 277-8581, Japan}

\begin{abstract}
We report results of nuclear magnetic resonance (NMR) experiments in
SrCu$_{2}$(BO$_{3}$)$_{2} $, a quasi two-dimensional spin system with a singlet
ground state. When magnetic field is applied along the $c$-axis, each of the 
quadrupole split Cu resonance lines splits further into four lines. 
The spin-echo intensity for some of the split lines oscillates
against the separation time between $\pi$/2 and $\pi$ rf-pulses.  These
phenomena are due to strong nuclear spin-spin coupling mediated by the 
electronic spin system, which exists only within a pair of nuclei. 
Thus the results provides direct evidence for the
dimer singlet gorund state in this material.
\end{abstract}



\begin{figure}[tbp]
\centering
\includegraphics[width=13cm,height=7cm, keepaspectratio,clip]{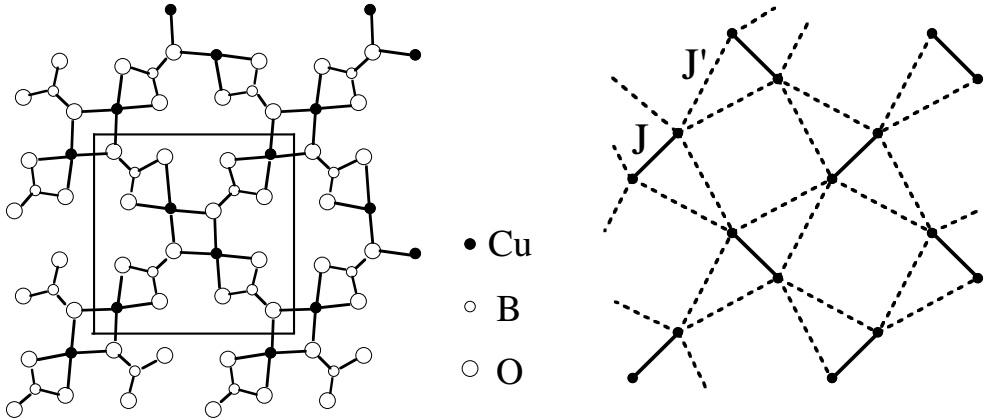}
\caption{Structure of the magnetic layer in SrCu$_2$(BO$_3$)$_2$ (left panel) and 
the Shastry-Sutherland model (right panel).}
\label{structure}
\end{figure}

The discovery of excitation gap and quantaized magnetization plateaus in the
quasi-two-dimensional spin system SrCu$_2$(BO$_3$)$_2$ by Kageyama {\it et
al.}\cite{kageyama991} have since stimulated vast amount of experimental and 
theoretical work. As shown in \fref{structure}, the Cu$^{2+}$ spins ($s$=1/2) 
in this compound form a planar network of dimers, which is
identical to the Shastry-Sutherland (SS) model\cite{shastry811} when only
the nearest neighbor intra-dimer exchange ($J$) and the second nearest 
inter-dimer exchange ($J'$) interactions are retained.

Extensive studies on the SS-model have revealed interesting novel 
aspects~[3-11].
First, the simple direct-product of dimer singlet is an exact eigenstate for
any value of $J'/J$ and is the ground state when $J'/J$ is smaller than a
certain critical value~\cite{shastry811,miyahara991}. In the opposite limits, 
$J'/J \gg 1$, the model reduces to the nearest neighbor Heisenberg model 
on a square-lattice with the obvious Neel order.
It was proposed that the two phases are separated by a first order transition 
at $T=0$ near $J'/J=0.7$~\cite{miyahara991}.
For SrCu$_2$(BO$_3$)$_2$, the exponential decrease of susceptibility and
Cu nuclear spin-lattice relaxation rate at low temperatures\cite{kageyama991} 
indicate that the ground state is singlet with a finite energy gap for excited states.
The magnitude of the excitation gap $\Delta$ is determined to be 35K from
the inelastic neutron scattering\cite{kageyama001} and the electron spin 
resonance\cite{nojiri991} experiments.

Secondly, excitations from the dimer singlet ground state in the SS-model
have extremely localized character.  Using perturbation expansion in $J'/J$,
Miyahara {\it et al.} found that hopping of an excited triplet from
one dimer to another is allowed only from sixth order\cite{miyahara991}.
A very small dispersion width of 0.2 meV for the magnetic excitations was indeed
observed by the neutron inelastic scattering experiments in 
SrCu$_2$(BO$_3$)$_2$\cite{kageyama001}. The most striking feature
of SrCu$_2$(BO$_3$)$_2$ is the plateaues in the magnetization curve under
high magnetic field at fractional values (1/8, 1/4, and 1/3) of the fully saturated
magnetization\cite{kageyama991,onizuka001}.
It is proposed that small kinetic energy of the triplets and weak repulsive
interaction between them are responsible for the formation of superstructure of 
triplet dimers at these plateaus\cite{miyahara001,momoi001},
although no direct experimental evidence has been obtained yet.

The nature of quantum phase transition in the SS-model, however, is not well
understood yet. Possibilities have been discussed for intermediate phases 
between the dimer singlet and the Neel state for a certain range of $J'/J$ 
such as a plaquette-type singlet state\cite{koga001} or a helical spin order\cite{mila961}. 
The exchange parameters appropriate for SrCu$_2$(BO$_3$)$_2$ were estimated
as $J=85$~K and $J' =54$~K ($J'/J=0.64$) by fitting the magnetic
susceptibility data to numerical simulation of the SS-model\cite{miyahara001}.  
Analysis of the energy of various excitation modes has lead to slightly different values, 
$J=72$~K and $J'=43$~K ($J'/J=0.60$)\cite{knetter001}.  For such parameter values,
theories appear to agree that the ground state is the dimer singlet.
Various properties of SrCu$_2$(BO$_3$)$_2$ known so far are indeed compatible with
the dimer singlet ground state, although it may be at close proximity to a quantum 
critical transition.

In this letter, we report the results of nuclear magnetic resonance (NMR)
experiments at $^{63}$Cu and $^{11}$B nuclei (both with nuclear spin 3/2) in
SrCu$_2$(BO$_3$)$_2$. We observed four-fold splitting of Cu NMR lines 
and sinusoidal oscillation of Cu spin echo intensity against the separation 
time between the $\pi$/2 and $\pi$ rf-pulse.  These phenomena are natually 
explained by strong nuclear spin-spin coupling acting only within a pair, 
therefore, provide direct evidence for the dimer singlet ground state.

The NMR measurements were performed on a single crystal prepared by the
traveling-solvent-floating-zone method and cut into an approximately cubic 
shape (2.1x2.3x1.9 mm).  The applied magnetic field $H$ did not exceed 8~T, 
therefore, the Zeeman energy was much smaller than the excitation gap. 
The NMR spectra were obtained from the Fourier transform of the spin-echo signal.
SrCu$_2$(BO$_3$)$_2$ has tetragonal structure with $I\overline{4}2m$ space
group, in which Cu(BO$_{3}$) and Sr layers stack alternately\cite{smith911}.  
The atoms in the magnetic Cu(BO$_{3}$) layer depicted in Fig. 1 are not strictly 
coplanar and the only symmetry operation at either Cu or B sites is the mirror 
reflection normal to $\langle 110 \rangle$.  Then the direction normal 
to the mirror plane coincides with one of the principal axes of the electric field 
gradient (EFG) and the magnetic hyperfine shift ($K$) tensors, however, both 
do not possess axial symmetry.  When the magneic field $H$ is along the $c$-axis, 
the quadrupolar and the magnetic shifts are identical for all Cu (B) sites but the 
$c$-axis is not a principal axis of the EFG or $K$ tensors.  For other field 
directions, there are more than two inequivalent sites.  For example, there are 
two Cu (B) sites for $H \parallel [110]$.  At one of these sites, $H$ is normal
to the mirror plane and thus along one of the principal axes of EFG and $K$
tensors.

\begin{figure}[tbp]
  \begin{minipage}{0.5\linewidth}
\includegraphics[scale=0.35]{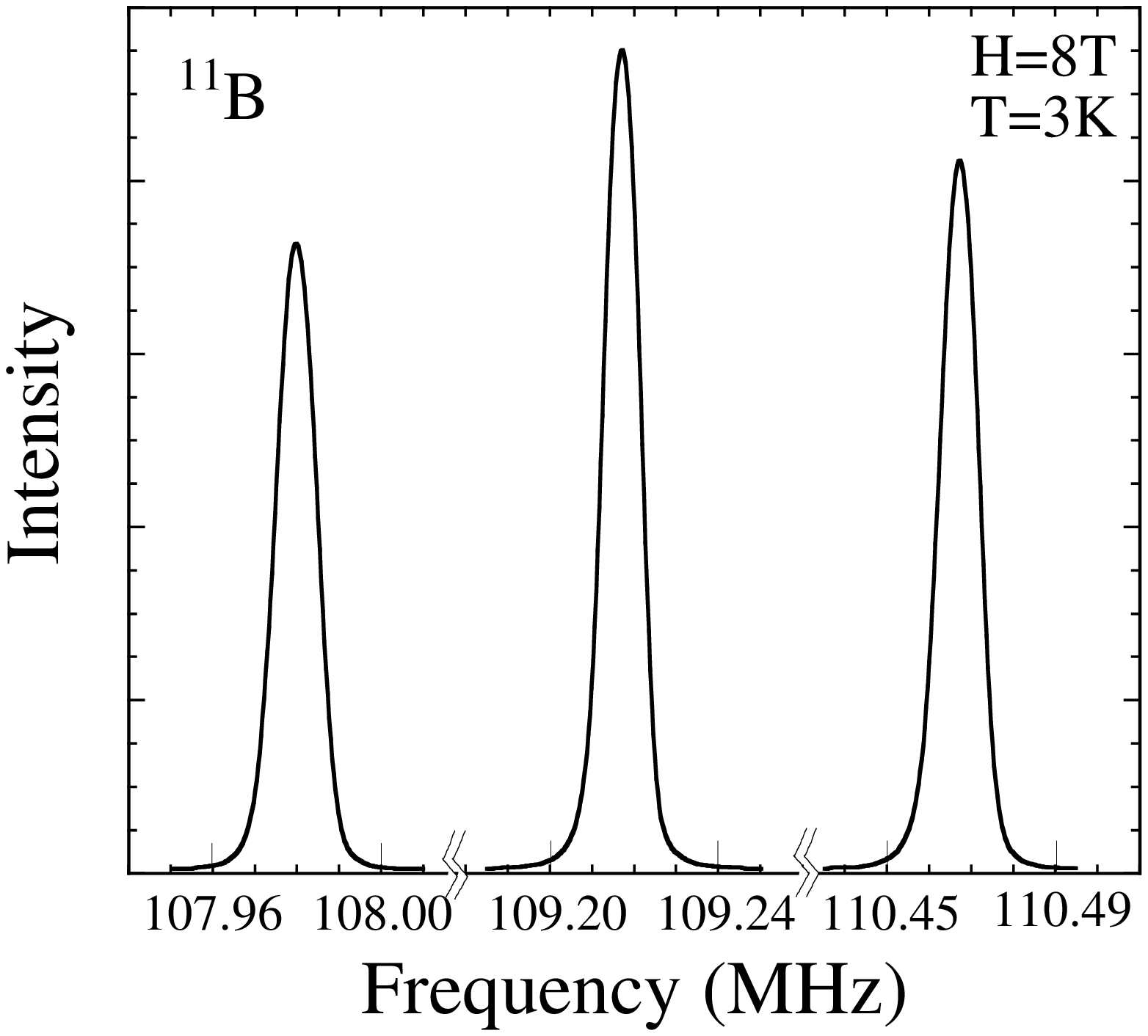}
\caption{The $^{11}$B spectrum at $T=$3K.  The magnetic field is 8 T parallel to the $c$-axis.}
\label{Bspectrum}
\end{minipage}
\begin{minipage}{0.5\linewidth}
\includegraphics[scale=0.35]{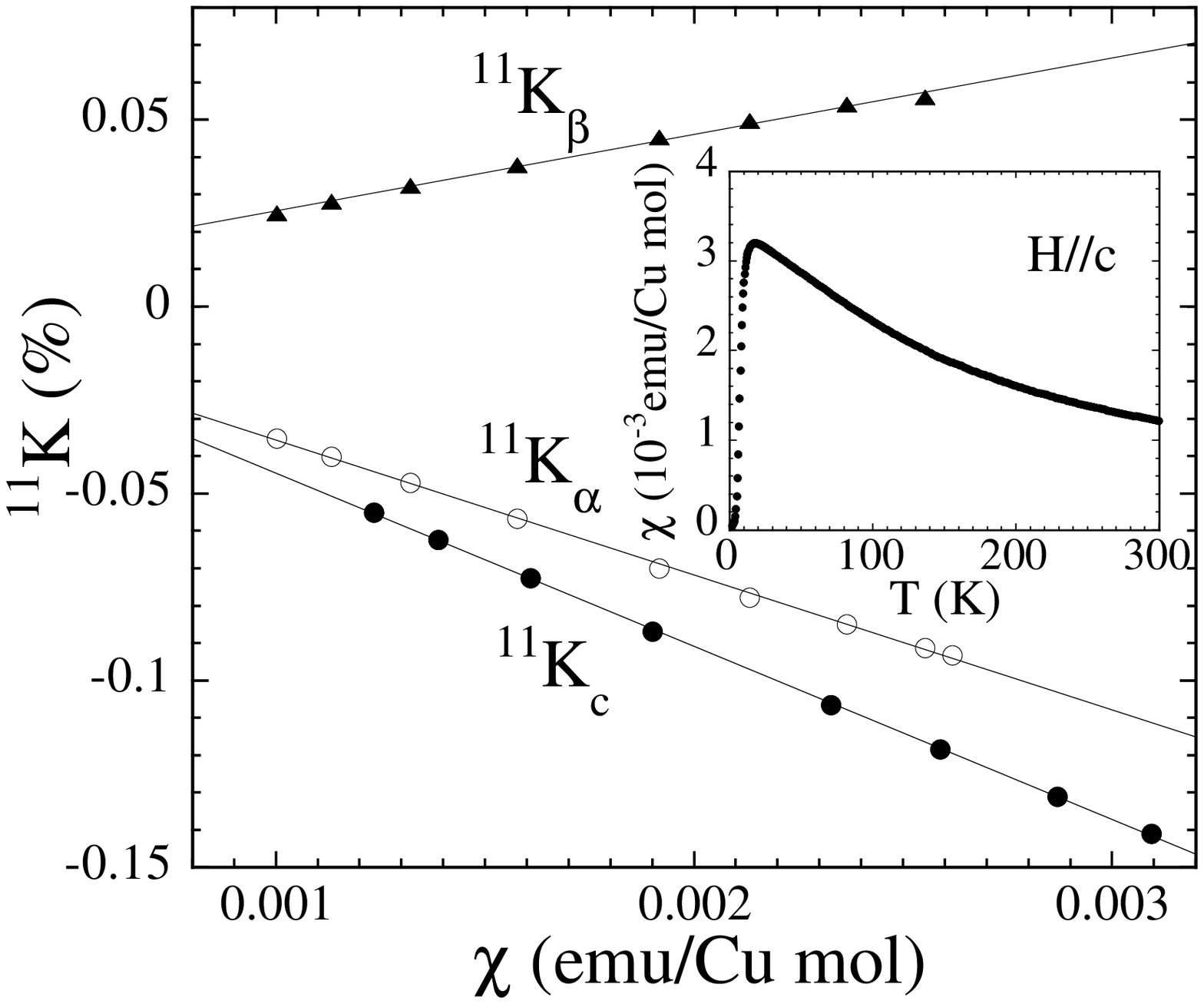}
\caption{The shifts at B sites are plotted against the 
susceptibility.  The solid lines show fitting to linear relation.}
\label{bkchi}
\end{minipage}
\end{figure}

\Fref{Bspectrum} shows the $^{11}$B NMR spectrum at $T=3$~K for
$H=8$~T along the $c$-axis.  The spectrum consists of three lines split by 
electric quadrupole interaction, yielding the quadrupole splitting
$^{11}\nu_{cc} = ^{11}V_{cc}e^{11}Q/2h = 1.25$~MHz, where $^{11}V_{cc}$ is
the $cc$-component of the EFG tensor and $^{11}Q$ is nuclear quadrupole moment. 
Other components of quadrupole splitting are obtained from the spectrum for 
$H\parallel [110]$ as $^{11}\nu_{\alpha\alpha}=0.694$~MHz, and
$^{11}\nu_{\beta\beta}=0.555$~MHz, where $\alpha$ is a principal axis and
$\beta$ is perpendiculat to it.

The magnetic hyperfine shift $^{11}K$ at B sites along three directions $c$,
$\alpha$ and $\beta$ are determined from the position of the central line for the
transition $I_z=1/2 \leftrightarrow -1/2$ at $H=8$~T.
There are good linear relations between the shifts and the magnetic
susceptibility $\chi$ as shown in \fref{bkchi}.  From the slope of the $K vs. \chi$ plot, 
three components of the hyperfine coupling tensor defined as 
$^{11}A_{ii}=^{11}K_{ii}/(\mu_{B}N_{A}\chi_{ii})$ are determined as $^{11}A_{cc}=-0.259$,
$^{11}A_{\alpha\alpha}=-0.202$, and $^{11}A_{\beta\beta}=0.115$ in unit of
T/$\mu_{B}$.  The coupling constant $^{11}A_{ii}$ represents the $i$-component of the
hyperfine field at B nuclei provided that each Cu had a uniform moment of 1~$\mu_B$ along the 
$i$-direction.  We found that the dipolar field from Cu spins alone does not
account for the anisotropy of $^{11}A_{ii}$, indicating sizable anisotropy
of the transferred hyperfine interaction.  The temperature dependence of $\chi_{cc}$ 
is shown in the inset of \fref{bkchi}.  Since the susceptibility was measured at 
1~T and the magnetization is not linear in field when the tempearture is much lower 
than the gap, only the data above the peak temperature of $\chi$ are used
in the $K vs. \chi$ plot. 

\begin{figure}[tbp]
\begin{minipage}{0.5\linewidth}
\includegraphics[scale=0.35]{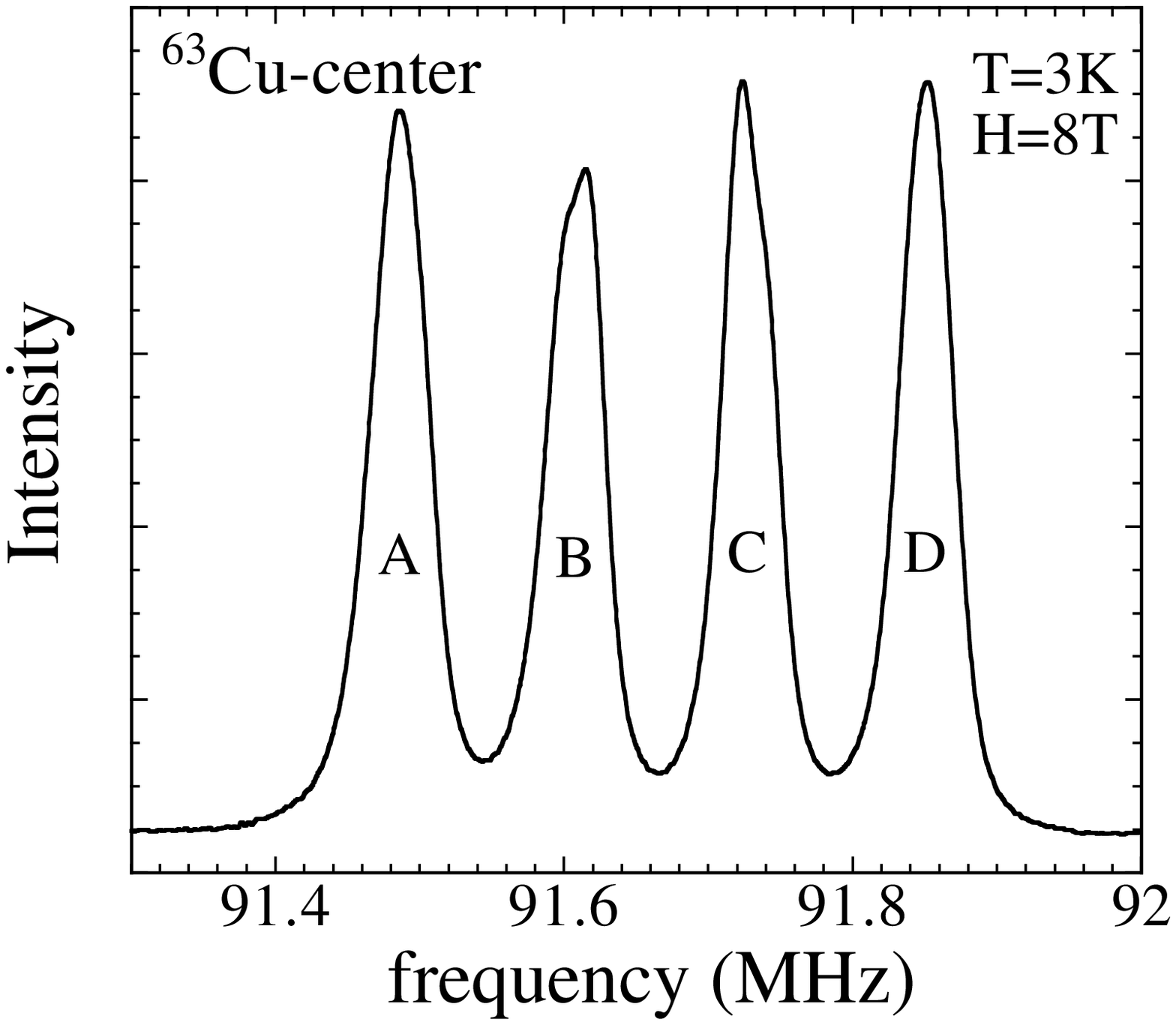}
\caption{The $^{63}$Cu NMR spectrum for central line at $T=$3 K.  The magnetic field is 8 T parallel to the $c$-axis.}
\label{cuspectrum}
\end{minipage}
\begin{minipage}{0.5\linewidth}
\includegraphics[scale=0.35]{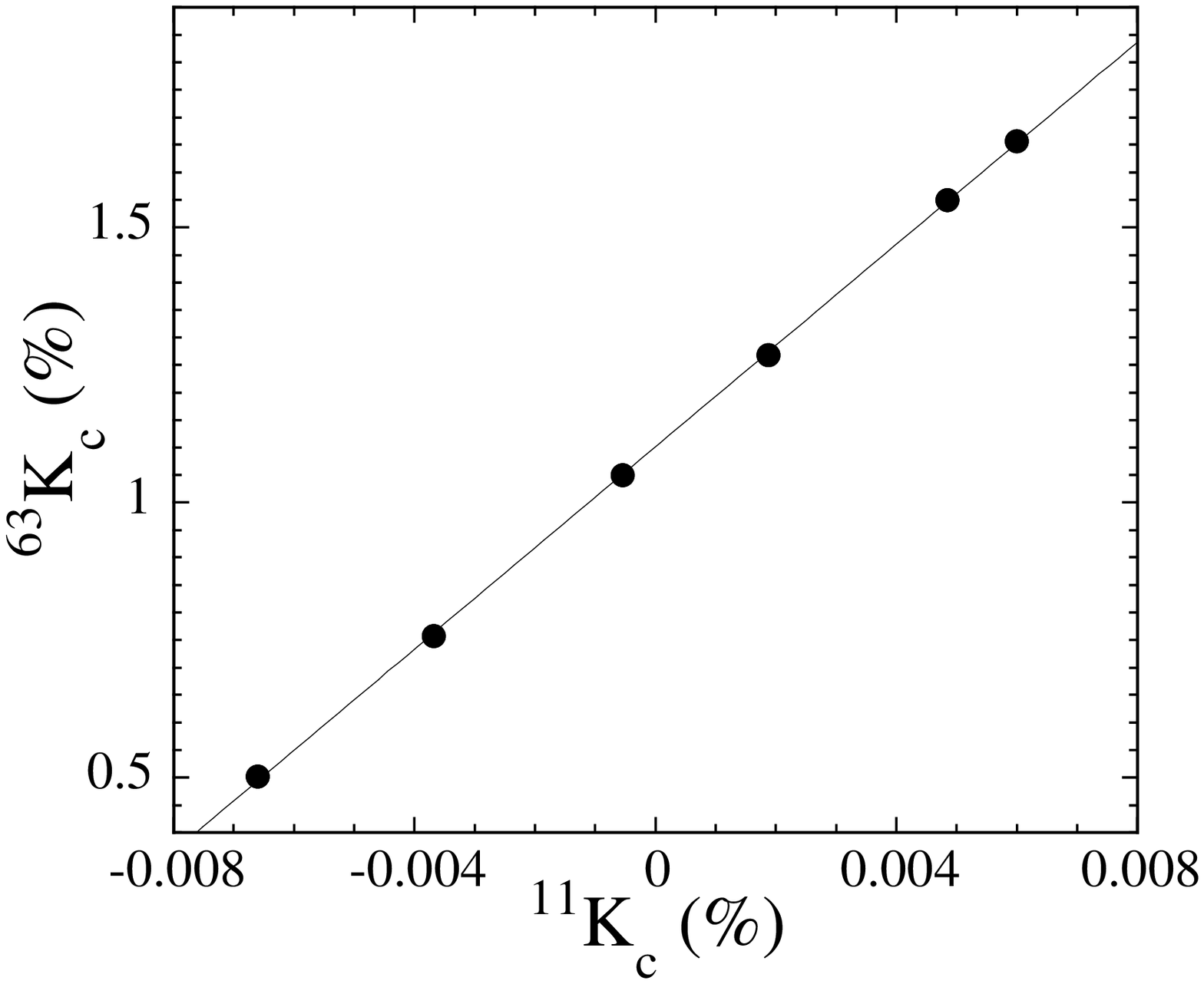}
\caption{The magnetic hyperfine shift at Cu sites ($^{63}K_{c}$) is plotted
against the shift at B sites ($^{11}K_{c}$).  Both were taken at the same field $H=8$~T along
the $c$-axis in the temperature range below 4.2~K.}
\label{kk}
\end{minipage}
\end{figure}

We now turn to the results at Cu sites.  \Fref{cuspectrum} shows the
$^{63}$Cu NMR spectrum for the central transition at $H=8$~T along the 
$c$-axis at $T=3.0$~K.  The spectrum consists of equally spaced four lines in 
spite of the fact that all Cu sites are equivalent for $H \parallel c$.
Similar four-peaks structure was observed for the quadrupole-split satellite
spectra.  The quadrupolar splitting for $^{63}$Cu nuclei was obtained as
$^{63}\nu_{cc}=22.13$~MHz.  As will be explained shortly, the four-fold splitting 
is due to the coupling of two Cu nuclear spins on the same dimer.  
The magnetic shift is then determined from the average frequency of the four peaks.
The observation of Cu NMR signal is limitted to $T \le 4.2$~K because of the
short spin-spin relaxation time at higher temperatures. 
The hyperfine coupling constant at Cu nuclei $^{63}A_{cc}$ was determined by plotting the
shift for $^{63}$Cu against the shift for $^{11}$B, both taken at the same field and 
several temperatures below 4.2~K as shown in \fref{kk}.  From the slope of this plot 
and the value of $^{11}A_{cc}$ determined above,
we obtain $^{63}A_{cc}=-23.76 \pm 0.15$~T/$\mu_{B}$.  Complete determination
of the EFG and $K$ tensors at B and Cu sites will be presented in a separate paper.

\begin{figure}[tbp]
\centering
\includegraphics[scale=0.4]{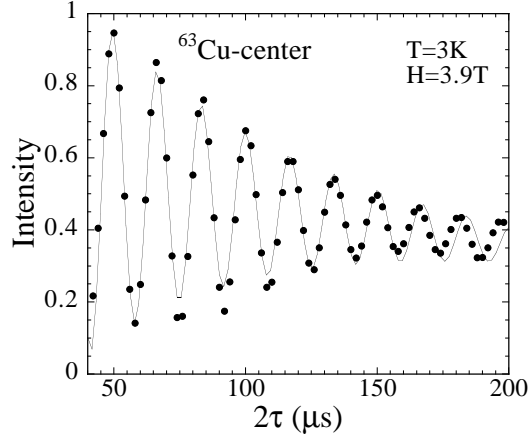}
\caption{The spin-echo intensity of the peak B in \fref{cuspectrum} for the
center line of $^{63}$Cu NMR spectrum is plotted as function of 2$\tau$,
where $\tau$ is the separation
time between $\pi/2$ and $\pi$ pulses. The data is taken at $H=3.9$~T along
the $c$-axis and $T=3$~K.}
\label{oscillation}
\end{figure}

We found remarkable oscillation of the spin-echo intensity of the $^{63}$Cu
resonance as shown in \fref{oscillation}.
Here the spin-echo intensity of the peak B of the central line is plotted
against twice the the separation time $\tau$ between $\pi$/2 and $\pi$ rf-pulses.
Such oscillation is observed only for the peaks B and C.  Similar
oscillation was observed for the satellite spectra, although at different peaks. 
For the high frequency sattelite, only the peaks C and D show the echo-oscillation, 
while for the low frequency satellite, oscillation was observed for the peaks A and B.

The line splitting and the echo-oscillation are both explained easily by 
Rudermann-Kittel-Kasuya-Yosida type indirect interaction between Cu nuclear
spins mediated by the electron spin system.
The indirect coupling between two nuclear spins $I_{1}$ and $I_{2}$
are generally expressed as\cite{takigawa941}
   \begin{eqnarray}
       H_{ind}=-a_{z}I_{1}^{z}I_{2}^{z} ,
   \label{iicoupling}
   \end{eqnarray}
   \begin{equation}
      a_{z}=\hbar^{2}\gamma_{N1} \gamma_{N2} (g_{z}A_{z})^2\chi_{12}
   \label{RKKY}
   \end{equation}
where $\gamma_{Ni}$ is the nuclear gyromagnetic ratio for $I_{i}$ and only
the on-site hyperfine coupling constant along the field direction $A_{z}$ is considered.  
The non-local electron spin susceptibiliry $\chi_{12}$
describes the spin polarization at site 1 when a fictitious magnetic
field were applied only to the spin at site 2.  At $T$=0 it is expressed as
   \begin{equation}
   \chi_{12} = \sum_{n} \frac{\langle 0 | S_{2z} | n \rangle \langle n|
S_{1z} | 0 \rangle} {E_{n}} + c.c. ,
   \end{equation}
where $| 0 \rangle$ is the gronud state and $| n \rangle$ is an excitated
state with the energy $E_{n}$.  In case of an isolated dimer with the exchange $J$,
  \begin{equation}
     \chi_{12}=-\frac{1}{2J} .
     \label{dimer}
\end{equation}
\Eref{iicoupling} implies that local magnetic field acting on $I_{1}$ is
produced by $I_{2z}$, which can take four different values $\pm1/2$ or $\pm3/2$.  
This results in the four-fold
splitting of the NMR line for $I_{1}$.
Since $I_{2}$ can be either of the two isotopes $^{63}$Cu or $^{65}$Cu,
there should be strictly speaking
eight-fold splitting. However, both isotope have spin 3/2 and the difference
in $\gamma$ is only 7~\%.
Thus they are not experimentally resolved.  Assuming that the spectrum in
\fref{cuspectrum}
represent the isotopic averagy for $I_{2}$, the value of $a_{z}$ for the
case when $I_{2}$ is a $^{63}$Cu
nucleus is deduced as $a_{z}/h = 119$~kHz.

It is well known that such a nuclear spin coupling gives rise to spin-echo
oscillation~\cite{abragham}
\begin{equation}
     I(2\tau)=\cos(a_{z} \tau /\hbar)
\end{equation}
when the two spins are identical (like spins).  Because of the quarupole
splitting, only those $^{63}$Cu nuclei
with $I_{z}=\pm 1/2$ behave as like spins when the central line for the
transition
$I_{z} =1/2 \leftrightarrow -1/2$ is being observed.  This is why
only the peaks B and C show oscillation.  The selection rules for the
sattelite lines is explained
in a similar manner. The data in \fref{oscillation} is fit to a sum of
oscillatory and non-oscillatory parts both
decaying exponantially with different time constant.  The oscillation
frequency is obtained as
$a_{z} /h = 119$~kHz in agreement with the value deduced from the splitting
of the spectrum.

The four-fold splitting of the spectrum and the coherent spin-echo
oscillation indicate that
nuclear spins are strongly coupled in pairs and the couplings between pairs
are very weak.  The results thus provide direct evidence for the dimer-singlet 
ground state in SrCu$_{2}$(BO$_{3}$)$_{2}$.  If one nucleus were coupled to many 
neighbors, for example as one would expect for the plaquette
type singlet state, the spectrum should consist of many lines, which with
random isotopic configuration would result in a single broad peak.  Likewise, 
random superposition of echo-oscillation containing
many frequency components would show rapid damping. The long-lived coherent
echo-oscillation implies that the inter-dimer coupling if any is orders of magnitude smaller
than the intra-dimer coupling.  Similar phenomena have been observed also for the 
chain Cu sites in the ladder-chain composite material Sr$_{14}$Cu$_{24}$O$_{41}$, 
where dimer singlet state is formed due to charge order in the one-dimensional 
chain subsystem~\cite{takigawa981}.

For the case of an isolated dimer, the coupling constant $a_z/h=119$~kHz correspoonds
to the exchange constant $J=75$~K.  Here we have used equations (\ref{iicoupling}), 
(\ref{RKKY}) and (\ref{dimer}) with the values $A_{c}=-23.76$~T/$\mu_{B}$ and $g_{c}=2.28$, 
the latter being obtained from the ESR measurements~\cite{nojiri991}.  
This value of $J$ is remarkablly close to the intra-dimer exchange estimated for
SrCu$_{2}$(BO$_{3}$)$_{2}$.  Although $J'/J$ is not a small number for 
SrCu$_{2}$(BO$_{3}$)$_{2}$, the spin correlation 
within a dimer must remain almost unchanged from an isolated dimer.

In conclusion, $^{63}$Cu and B NMR measurement has been performed on a
single crystal of SrCu$_{2}$(BO$_{3}$)$_{2}$.  The values of the hyperfine coupling constant
and the quadrupole splitting were determined.  The spectrum of Cu splits into
four lines and the spin-echo intensity oscillates against $2\tau$ with the
oscillation frequency equal to the line splitting.  
They are caused by the nuclear spin-spin coupling
mediated by the strong intradimer coupling of the electron spins.  These results
provide firm evidence that the ground state is indeed the dimer singlet state.

\ack
The work was supported by the Grant-in Aid for Scientific Research No.
10304027 for the Japan Society of the Promotion of Science and the Priority Area (A)
on "Novel quantum phenomena in transitioin metal oxides" and the Priority
Area (B) on "Field-induced new quantum phenomena in magnetic systems" from
the Ministry of Education, Calture, Sports Science and Technology of Japan.


\section*{References}
\begin{thebibliography}{99}
\bibitem{kageyama991} Kageyama H, Yoshimura K, Stern R, Mushnikov N V,
Kato M, Kosuge K, Slichter C P, Goto T and Ueda Y 1999 {\it Phys. Rev.
Lett.} {\bf 82} 3168
\bibitem{shastry811} Shastry B S and Sutherland B 1981 {\it Physica} B {\bf
108} 1069
\bibitem{mila961} Mila F and Albrecht M 1996 {\it Europhys. Lett.} {\bf 34}
145
\bibitem{miyahara991} Miyahara S and Ueda K 1999 {\it Phys. Rev. Lett.} {\bf
82} 3701
\bibitem{weihong991} Weihong Z, Hamer C J and Oittma J 1999 {\it Phys. Rev.}
B{\bf 60} 6608
\bibitem{miyahara001} Miyahara S and Ueda K 2000 {\it Phys. Rev.} B{\bf 61}
3417
\bibitem{momoi001} Momoi T and Totsuka K 2000 {\it Phys. Rev.} B{\bf 61}
3231, B{\bf 62} 15067
\bibitem{miyahara002} Miyahara S and Ueda K 2000 {\it J. Phys. Soc. Jpn.}
{\bf 69} Suppl. B. 72
\bibitem{muller001} M\"{u}ller-Hartmann E, Singh R R P, Knetter C and Uhrig
G S 2000 {\it Phys. Rev. Lett.} {\bf 84} 1808
\bibitem{koga001} Koga A and Kawakami N 2000 {\it Phys. Rev. Lett.} {\bf 84}
4461
\bibitem{knetter001} Knetter C, B\"{u}hler A, M\"{u}ller-Hartmann E and
Uhrig G S 2000 {\it Phys. Rev. Lett.} {\bf 85} 3958
\bibitem{kageyama001} Kageyama H, Nishi M, Aso N, Onizuka K, Yoshihama T,
Nukui K, Kodama K, Kakurai K and Ueda  Y 2000 {\it Phys. Rev. Lett.} {\bf
84} 5876
\bibitem{nojiri991} Nojiri H, Kageyama H, Onizuka K, Ueda Y and Motokawa M
1999 {\it J. Phys. Soc. Jpn.} {\bf 68} 2906
\bibitem{onizuka001} Onizuka K, Kageyama H, Ueda Y, Goto T, Narumi Y and
Kindo K 2000 {\it J. Phys. Soc. Jpn.} {\bf 69} 1016
\bibitem{smith911} Smith R W and Keszler D A 1991 {\it J. Solid State Chem.}
{\bf 93} 430
\bibitem{takigawa941} Takigawa M 1994 {\it Phys. Rev.} B{\bf 49} 4158
\bibitem{abragham} Abragham A 1961 {\it The Princiles of Nuclear Magnetism}
(Oxford: Oxford
University Press) p~497
\bibitem{takigawa981} Takigawa M, Motoyama N, Eisaki H and Uchida S 1998
{\it Phys. Rev.}
1998 B{\bf 57} 1124
\end{thebibliography}
\end{document}